\documentclass[twocolumn,showpacs,amsmath]{revtex4}
\usepackage{graphicx,amsmath,amsfonts, amssymb}
\newcommand{\Journal}[4]{#1 \textbf{#2}, #3 (#4)}

\newcommand{\FeVAl}{Fe$_2$VAl}

\newcommand{\BiTe}{Bi$_2$Te$_3$}

\begin{document}

\title{Electronic and thermoelectric properties of Fe$_2$VAl: The role of defects and disorder}

\author{Daniel I. Bilc}
\author{Philippe Ghosez}
\affiliation{Physique Th\'eorique des Mat\'eriaux, Universit\'e de Li\`ege (B5), B-4000 Li\`ege, Belgium}


\begin{abstract}

Using first-principles calculations, we show  that \FeVAl\ is an indirect band gap semiconductor.  Our calculations reveal that its, sometimes assigned, semimetallic character is not an intrinsic property but originates from the antisite defects and site disorder, which introduce localized ingap and resonant states changing the electronic properties close to band gap. These states negatively affect the thermopower $S$ and power factor PF=$S^2\sigma$, decreasing the good thermoelectric performance of intrinsic \FeVAl.     

\end{abstract}

\maketitle

Energy-related issues are becoming more and more crucial. Devices based on thermoelectric (TE) materials are very appealing in this context, and they are one of the main thrusts in energy research on the global scale. They can be used for cooling or heating  and for energy generation from recovered waste heat. The efficiency of a TE material depends on the dimensionless TE figure of merit, ZT=$(S^2\sigma$T$)/\kappa_{th}$, where $\sigma$ is the electrical conductivity, $S$ is the thermopower, T is the absolute temperature,  and $\kappa_{th}$ is total thermal conductivity which has electronic and lattice contributions. Improving the TE efficiency is not obvious because the different parameters entering ZT are linked and compete with each others. Moreover, ZT has to be maximized in the regime at which the TE device will be operated and other aspects must also be considered, such as the cost of materials, their toxicity and availability. For many applications, large scale installations will be essential for harnessing the full potential of thermoelectricity and the use of less efficient but cheap compounds might offer a valuable solution.

In this respect, \FeVAl\ has electronic properties potentially interesting for TE applications (ZT$\sim$0.13 at 300K for Fe$_2$VAl$_{0.9}$Ge$_{0.1}$~\cite{Nishino2006} and ZT$\sim$0.15 at 300K for Fe$_2$VAl$_{0.9}$Si$_{0.07}$Sb$_{0.03}$~\cite{Mikami2009}). \FeVAl\ has a L2$_1$ Heusler structure. It is more than a decade since Nishino {\it et al.} have reported its unusual properties, but the ground state of the intrinsic system is still unclear~\cite{Nishino1997}. \FeVAl\ shows a semiconducting or semimetallic behaviour with a pseudogap of $\sim$0.1-0.27 eV~\cite{Lue1998, Okamura2000}. \FeVAl\ is nonmagnetic (no long range FM order) but possesses magnetic anti-site defects and superparamagnetic clusters~\cite{Vasundhara2008b, Lue2001}. It has a large specific heat at low T and first it was suggested to be a possible candidate for a 3d heavy-fermion system~\cite{Nishino1997, Nishino2000}. Later, field-dependent specific heat measurements showed that the large specific heat was not an intrinsic behaviour and it was assigned to magnetic defects~\cite{Lue1999}. \FeVAl\ has also a negative resistivity slope at high T~\cite{Nishino2000}. It has a dominant p-type transport character, with a high hole concentration of $n_h$=4.8x10$^{20}$ cm$^{-3}$~\cite{Kato1998} and shows a very large residual resistivity~\cite{Nishino1997}. All these properties suggest that defects and disorder play an important role in this material. 

Previous electronic structure calculations on  \FeVAl\ were based on density functional theory (DFT) using usual exchange-correlations functionals such as the generalized gradient and local density approximations (GGA, LDA) and predicted \FeVAl\ to be a compensated semimetal with a deep pseudogap of $\sim$0.1-0.2 eV~\cite{Weht1998, Singh1998, Guo1998}. The presence of pseudogap with a finite density of states at the Fermi level is supported by optical reflectivity~\cite{Okamura2000} and NMR~\cite{Lue1998} experiments, but the compensated character of carriers is in contradiction with Hall measurements~\cite{Kato1998} which found excess of holes. Usual DFT functionals are known to underestimate semiconductor band gaps and usually fail to describe strongly correlated systems. Hybrid functionals often allow to circumvents these problems and so constitute a promising alternative to better characterize \FeVAl.  

In this letter we report the electronic and transport properties of \FeVAl\ using both the usual GGA functional of Perdew, Burke, and Ernzerhof (PBE~\cite{PBE}) and the recently developed B1-WC hybrid functional~\cite{Bilc2008}, which mixes the GGA of Wu and Cohen~\cite{WCGGA} with 16$\%$ of exact exchange. Whereas PBE reproduces the semimetallic character found in previous calculations, we predict \FeVAl\ to be a narrow gap semiconductor within B1-WC. Going beyond previous studies, we also investigate the role of single antisite defects and disorder on the transport properties. Our calculations strongly support that \FeVAl\ is intrinsically a semiconductor and that its semimetallic character originates from antisite defects and disorder, which introduce localized ingap and resonant states in the vicinity of the band gap. We show that such states negatively affect the power factor PF=$S^2\sigma$.

The calculations~\cite{DetailsAux} were performed using the augmented plane wave and local orbital (APW+lo) method as implemented in WIEN2k~\cite{WIEN2k}. We use the experimental lattice constant of 5.76 $\AA$ for a better comparison with previous calculations. The spin-orbit interactions and scalar relativistic effects were included. For the antisite defects and disordered configuration we considered supercells (SCs) with rhombohedral symmetry having 32 formula units (f.u) of \FeVAl\ which were derived from a 2x2x2 fcc cell with 4 atoms/cell along [1,1,1] direction~\cite{DetailsAux}. The transport calculations were performed using Boltzmann transport formalism within the constant relaxation time approximation using BoltzTraP~\cite{BoltzTraP}.

\begin{figure}[t]
 \centering\includegraphics[scale=0.40]{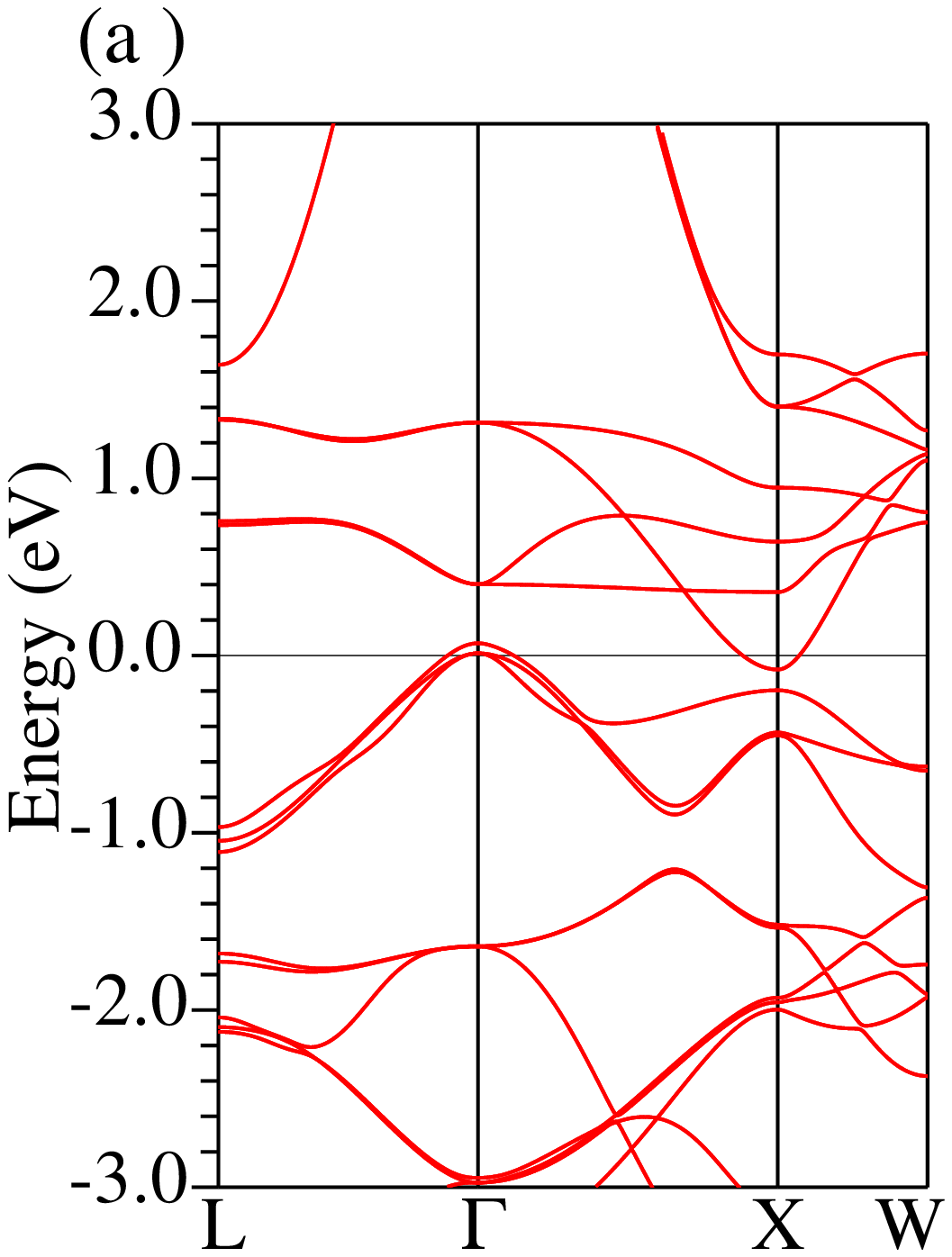}%
 \centering\includegraphics[scale=0.40]{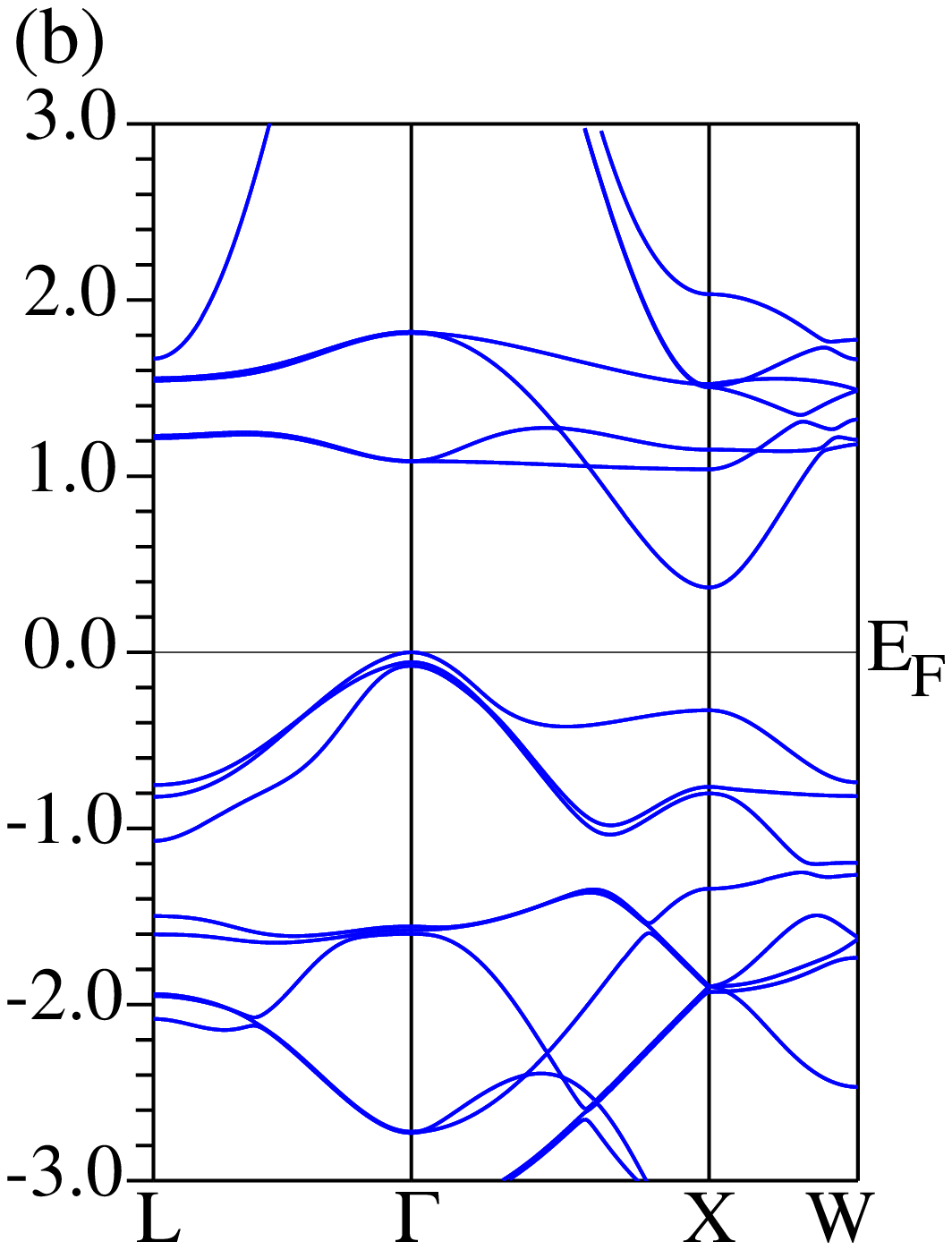}\\[-10pt]
 \caption{\label{bandfig} (Color online) Electronic band structure of fcc \FeVAl\ within: (a)PBE and (b)B1-WC.}
\end{figure}

First, we consider the electronic properties. Our calculations describe \FeVAl\ to be nonmagnetic within both PBE and B1-WC. Although, it is described as a semimetal within PBE (Fig.~\ref{bandfig}a), as consistently obtained in previous calculations, it is predicted to be a semiconductor with an indirect band gap E$_g$=0.34 eV (Fig.~\ref{bandfig}b), within B1-WC. This is not specific to B1-WC: other hybrid functionals, such as B3PW91, similarly predict a semiconducting character, although the exact value of the gap depends on the percentage of exact exchange included in the functional. In comparison to PBE, B1-WC opens an indirect band gap by shifting up in energy the lowest conduction band (CB) states with mixed V and Fe e$_g$ character with respect to the top valence band states with Fe t$_{2g}$ character. The states close to CB minimum (X point in Brillouin zone) have a highly dispersive V e$_g$ character, a very desired feature for good TE performance. We notice also that the charge of Al resulting from Bader analysis is equal to 1.68 $|e|$. This highlights that Al p states hybridize with V and Fe d states and that Al does not donate all its 3 electrons to the Fe-V network, as sometimes assumed~\cite{Weht1998}. 

\begin{figure}[t]
 \centering\includegraphics[scale=0.3, angle=0]{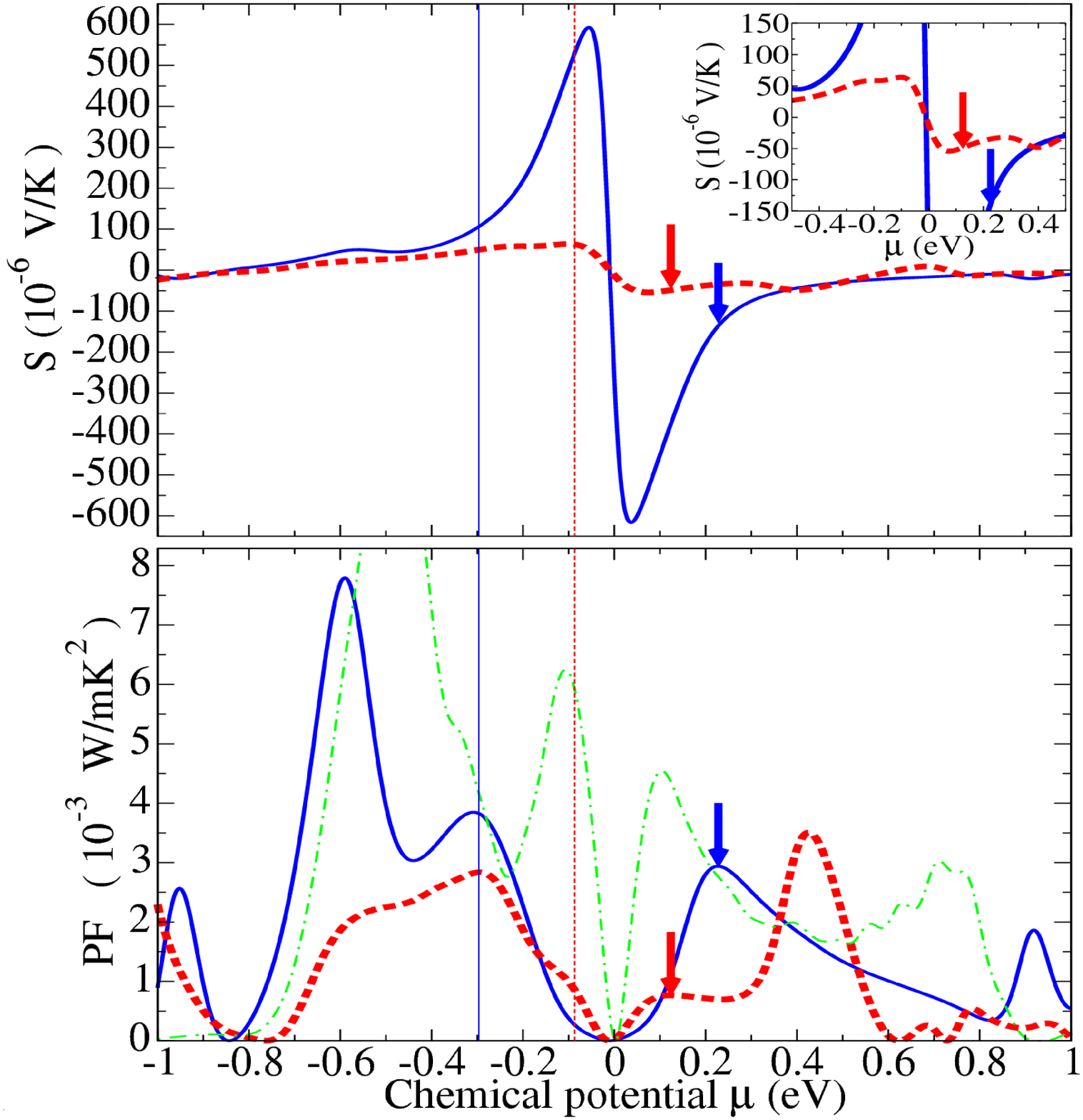}\\[-10pt]
 \caption{\label{SPFfccfig} (Color online) Thermopower $S$ ($S_{xx}$) and power factor PF=$S^2\sigma$ (PF$_{xx}$) as a function of chemical potential $\mu$ of fcc \FeVAl\ within B1-WC(continuous) and PBE(dashed) at 300 K using $\tau_e^{B1-WC}$ and $\tau_e^{PBE}$, respectively. $\mu$ corresponding to $n_h$=4.8x10$^{20}$ cm$^{-3}$ are indicated by vertical lines. The maximum PF and corresponding $S$ for n-type doping are indicated by arrows. PF of \BiTe\ within PBE using $\tau_e$=2.2x10$^{-14}$ s is included for comparison (dot-dashed).}
\end{figure}

Let us now focus on the transport properties, using the constant relaxation time approximation. Within this approximation, $S$ is independent of the relaxation time $\tau$, whereas $\sigma$ and the power factor PF=$S^2\sigma$ depend linearly on $\tau$. Mainly due to Al deficiency~\cite{Maksimov2001}, \FeVAl\ naturally forms as a hole-doped system with a carrier concentration $n_h$=4.8x10$^{20}$ cm$^{-3}$. The value of $\tau$ was estimated at this carrier concentration by fitting the electrical resistivity $\rho$ at 300 K to the experimental value of 0.75 m$\Omega$cm~\cite{Nishino2006, Vasundhara2008}. This yielded a hole relaxation time $\tau_h^{PBE}$=0.9x10$^{-14}$ s within PBE and $\tau_h^{B1-WC}$=1.4x10$^{-14}$ s within B1-WC. The electronic specific heat was also estimated for this hole-doped system .  We obtained a value of 1.00 mJ/molK$^2$ within B1-WC, in better agreement with the experimental estimate of 1.5$\pm$0.3 mJ/molK$^2$~\cite{Lue1999} than the PBE value of 0.76 mJ/molK$^2$.  

Electron doping of \FeVAl\ can be achieved from atomic substitution at Al site. We so estimated $\tau_e$ for electron doped Fe$_2$VAl$_{1-x}$M$_x$ (M=Si, Ge) systems by fitting $\rho$ at 300K to the experimental value of 0.65 m$\Omega$cm corresponding to a doping $x=0.03$~\cite{Nishino2006, Vasundhara2008}. Assuming that each atom M brings one additional electron, this corresponds to an electron concentration $n_e \sim$6.0x10$^{20}$ cm$^{-3}$, which adds to the initial $n_h$, assumed to be unchanged. Taking this into account, we get the electron relaxation times $\tau_e^{PBE}$=1.5x10$^{-14}$ s and $\tau_e^{B1-WC}$=3.4x10$^{-14}$ s. These values are slightly larger than those of the naturally formed hole-doped system, in qualitative agreement with the observation that the residual resistivity of \FeVAl\ decreases with doping at Al site~\cite{Nishino2006, Vasundhara2008}. 

In Figure~\ref{SPFfccfig}, we report the thermopower $S$ and power factor PF of electron-doped \FeVAl\ along the x-axis ($S$=$S_{xx}$, PF=PF$_{xx}$)  as a function of the chemical potential $\mu$. The amplitudes of $S$ and PF corresponding to $\mu$ for which the n-type PF reaches its maximum value are indicated by arrows. This is obtained for $\mu$=0.23 (resp. 0.12) within B1-WC (resp. PBE) and corresponds to a doping concentration $x=0.03$ (resp. 0.05) for n-type Fe$_2$VAl$_{1-x}$M$_x$ systems. Since we do not know the Al deficiency of these systems, we compare our values with  the experimental values for which the maximum PF were achieved ($ S \sim$ -120:-130 $\mu$V/K and  PF $\sim$4-6 mW/mK$^2$ at 300 K ~\cite{Nishino2006, Vasundhara2008}). Within PBE, \FeVAl\ has a semi-metallic character and $S$ reaches a maximum value of $\sim$ -55 $\mu$V/K while PF saturates around $\sim$0.7 mW/mK$^2$ for accessible $x$ values. These values remain similar within a wide range of $\mu$ and cannot explain the much larger values reported experimentally. In contrast, within B1-WC which describes \FeVAl\ as a semiconductor , we get larger values $S \sim$ -137 $\mu$V/K and PF $\sim$3 mW/mK$^2$ in close agreement with experimental data.  This demonstrates that a better description of \FeVAl\ is obtained when properly accounting for its semiconductor nature as obtained within B1-WC. 

\begin{figure}[t]
\centering\includegraphics[scale=0.3, angle=-90]{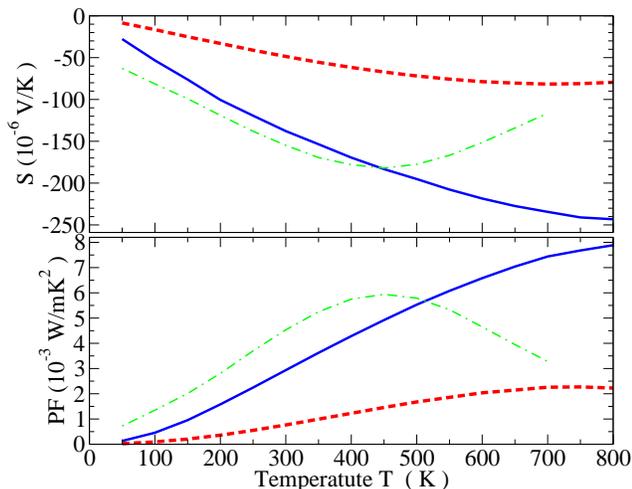}\\[-10pt]
\caption{\label{SPFvsTfccfig} (Color online) Temperature dependence of $S_{xx}$ and PF$_{xx}$ of fcc \FeVAl\ within B1-WC(continuous) and PBE(dashed) at $x$ values of 0.03 and 0.05, respectively. \BiTe\ values within PBE are also included (dot-dashed).}
\end{figure}

In Figure~\ref{SPFvsTfccfig},  we also report the temperature dependence of $S$ and PF at fixed concentrations, corresponding to $\mu$ for which the n-type PF reaches its maximum value within B1-WC and PBE at 300K (see Fig~\ref{SPFfccfig}). It can be seen that the theoretical $S$ and PF do not saturate even up to 800 K within B1-WC. This shows the very good potential of \FeVAl\ for TE performance at high temperatures (E$_g$=0.34 eV) in comparison with other thermoelectric materials like \BiTe\ (experimental E$_g$=0.15 eV). This contrasts however with the experimental observations, showing than $S$ and PF are saturating around 200-250 K. We infer that this deterioration of the TE performance comes from defects and disorder, which change the electronic properties near the band gap. 

In order to further prove this, we considered large SCs including 32 f.u. and performed B1-WC calculations including different types of single antisite defects and even considering a fully disordered configuration. Although the SC size corresponds to defect concentrations of $\sim 0.03$, typically one-order of magnitude larger than what is observed experimentally~\cite{Lue1999}, it is sufficient to treat the defects as isolated and to highlight their influence on the electronic and transport properties. 

\begin{figure}[t]
\centering\includegraphics[scale=0.3, angle=-90]{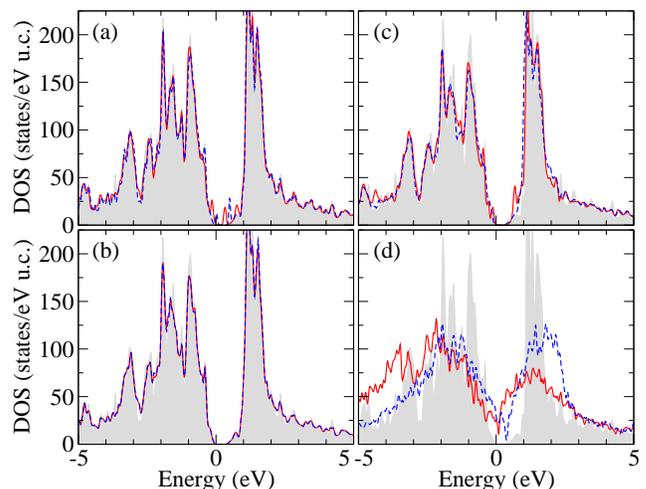}\\[-10pt]
\caption{\label{DOSfig} (Color online) Spin up(continuous) and down(dashed) total density of states (DOS) for: (a) V$_{Fe}$, (b) V$_{Al}$, (c) Fe$_{Al}$ antisite single defects, and (d) disordered configuration.  The DOS of intrinsic \FeVAl\ is shown as a gray background.} 
\end{figure}

As illustrated in Fig.~\ref{DOSfig}a, the V$_{Fe}$ antisite defect, arising from the permutation of one V atom with Fe, reduces the band gap of \FeVAl\ to $\sim$0.18 eV, by introducing localized d states into the gap, directly associated to the V and Fe atoms forming the defect. Moreover, this defect is magnetic with a magnetic moment of 4$\mu_B$/defect localized on the defect, a value which agrees well with that of 3.7$\mu_B$/defect found in specific heat and NMR experiments~\cite{Lue1999, Lue2001}. By contrast, the V$_{Al}$ antisite defect is non-magnetic and does not introduce any ingap state (Fig.~\ref{DOSfig}b). Finally , the Fe$_{Al}$ antisite defect introduces resonant Fe d states at the bottom (resp. top) of conduction (resp. valence) bands (Fig.~\ref{DOSfig}c). Again, this defect is magnetic with a magnetic moment of 4.6$\mu_B$/defect.  So, our B1-WC calculations reveal that some antisite defects are magnetic and that only those introduce localized ingap states and resonant states close to the gap region, significantly changing the electronic properties of \FeVAl.

In order to model further the effect of site disorder, we also considered a disordered configuration, arising from an arbitrary occupancy of the different sites within the SC and including 20 antisite defects (8 V$_{Fe}$, 6 V$_{Al}$, and 6 Fe$_{Al}$). The change in the electronic properties close to the gap region is even more obvious for this disordered configuration for which a semimetallic behaviour with a pseudogap and a magnetic moment of 53.5$\mu_B$/cell is obtained (see Fig.~\ref{DOSfig}d). These results clearly establish that the semimetallic character seen in experiments can be explained from antisite defects and disorder.       

\begin{figure}[t]
 \centering\includegraphics[scale=0.3, angle=-90]{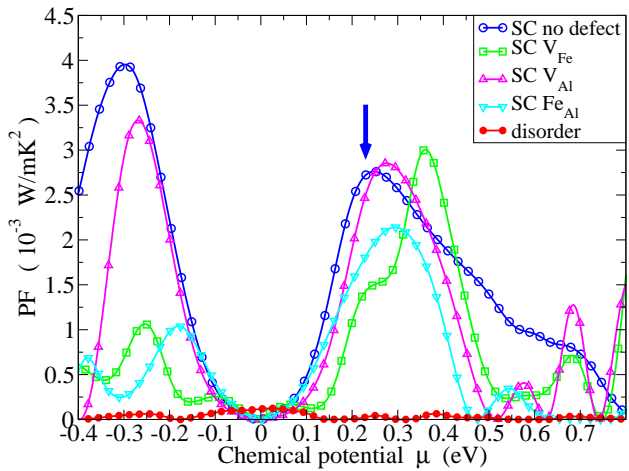}\\[-10pt]
 \caption{\label{PFAdeffig} (Color online) PF$_{xx}$ as a function of $\mu$ for the antisite defects and disordered configuration of \FeVAl\ within B1-WC at 300 K using $\tau_e^{B1-WC}$. $x$ value of 0.03 for the intrinsic \FeVAl\ is indicated by arrow. Note that for defects and disorder, this $x$ value is achieved at smaller $\mu$. } 
\end{figure}

It is now very interesting to explore the effect of the localized ingap and resonant d states on the transport properties. Mahan and Sofo have shown that a narrow energy (delta-shape) distribution of the electronic states participating in the electronic transport is needed in order to maximize ZT~\cite{Mahan1996}. Therefore, such localized ingap and resonant d states are expected to increase PF in the cases where these states have a significant weight with respect to the background states~\cite{Mahan1996}. However, as summarized in Fig.~\ref{PFAdeffig}, our transport calculations including antisite defects and disorder show that the localized ingap d states do not increase PF, which takes smaller or comparable values for accessible $n_e$ doping values. For defect concentration seen in experiment, the reduction should be less apparent but these calculations establish that antisite defects will never boost the TE performance.

\begin{figure}[b]
 \centering\includegraphics[scale=0.3, angle=-90]{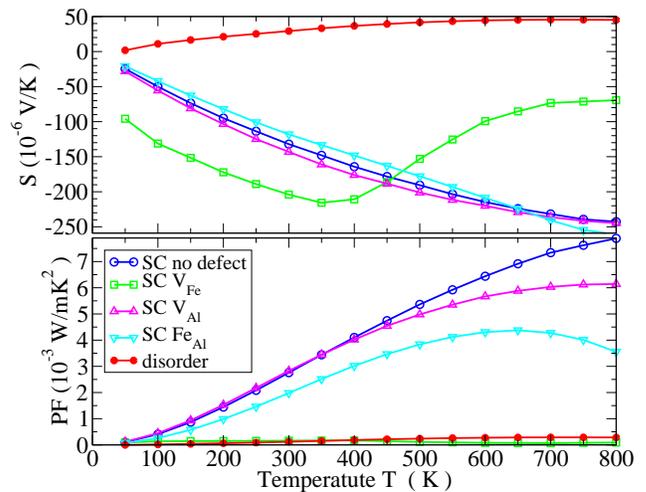}\\[-10pt]
 \caption{\label{SPFvsTdeffig} (Color online) Temperature dependence of $S_{xx}$ and PF$_{xx}$ for the antisite defects and disordered configuration of \FeVAl\ within B1-WC at $x$ value of 0.03. } 
\end{figure}

For $n_e$ value corresponding to the maximum PF of intrinsic \FeVAl\ (x=0.03), we show the temperature dependence of $S$ and PF in Figure~\ref{SPFvsTdeffig}. The antisite defects and disorder have a detrimental effect on PF, decreasing and saturating its values with T. For V$_{Fe}$ defect, $S$ and PF values are saturating at $\sim$350 K, behaviour seen in experiment. It is interesting to note that our disordered configuration have a ''hole-like'' dominated $S$, even at x=0.03. This suggests that the p-type character of \FeVAl\ may originate also partly from site disorder, and not only from off-stoichiometry of the constituents. 

In summary, our B1-WC calculations show that \FeVAl\ is an indirect narrow band gap semiconductor with a highly-dispersive V e$_g$ CB and three fold degenerate CB minimum, all features highly-compatible with good intrinsic TE performances. Our calculations including anti-site defects also demonstrate that the semimetallic character of \FeVAl\ seen in experiments can be explained from atomic disorder. Some anti-site defects are magnetic and introduce localized ingap and resonant states in the gap region. These defects tend to decrease the good intrinsic TE performances of \FeVAl.

\begin{acknowledgments}

We acknowledge financial support from Walloon Region through the CoGeTher, EnergyWall project.

\end{acknowledgments}

\end{document}